\documentclass{article}

\usepackage{PRIMEarxiv}

\usepackage[utf8]{inputenc} 
\usepackage[T1]{fontenc}    
\usepackage{hyperref}       
\usepackage{url}            
\usepackage{booktabs}       
\usepackage{amsfonts}       
\usepackage{amsmath,bm}
\usepackage{nicefrac}       
\usepackage{microtype}      
\usepackage{lipsum}
\usepackage{fancyhdr}       
\usepackage{graphicx}       
\usepackage{rotating,tabularx}
\usepackage{multirow}

\usepackage{bbm}
\usepackage{subcaption}
\usepackage{dsfont}

\usepackage{amsthm}
\usepackage[ruled,vlined]{algorithm2e}

\newcommand{\prob}{\operatorname{P}}
\newcommand{\vect}[1]{\boldsymbol{#1}}

\DeclareMathOperator*{\argmax}{arg\,max}

\graphicspath{{media/}}     

\title{A non-parametric estimator for Archimedean copulas
under flexible censoring scenarios and an application
to claims reserving
\thanks{\textit{\underline{Citation}}: 
\textbf{Michaelides, M., Cossette, H. \& Pigeon, M., A non-parametric estimator for Archimedean copulas
under flexible censoring scenarios and an application
to claims reserving (2024). Preprint.} 
}}

\author{
  Marie Michaelides\\
  Chaire Co-operators en analyse des risques actuariels \\
  Département de mathématiques \\
  Université du Québec à Montréal \\
  Montréal\\
  \texttt{michaelides.marie@uqam.ca} \\
  \And
  Mathieu Pigeon \\
  Chaire Co-operators en analyse des risques actuariels \\
  Département de mathématiques \\
  Université du Québec à Montréal \\
  Montréal\\
   \And
  Hélène Cossette \\
  École d'actuariat \\
  Université Laval \\
  Québec\\
}

\begin{document}
\maketitle

\begin{abstract}
With insurers benefiting from ever-larger amounts of data of increasing complexity, we explore a data-driven method to model dependence within multilevel claims in this paper. More specifically, we start from a non-parametric estimator for Archimedean copula generators introduced by Genest and Rivest (1993), and we extend it to diverse flexible censoring scenarios using techniques derived from survival analysis. We implement a graphical selection procedure for copulas that we validate using goodness-of-fit methods applied to complete, single-censored, and double-censored bivariate data. We illustrate the performance of our model with multiple simulation studies. We then apply our methodology to a recent Canadian automobile insurance dataset where we seek to model the dependence between the activation delays of correlated coverages. We show that our model performs quite well in selecting the best-fitted copula for the data at hand, especially when the dataset is large, and that the results can then be used as part of a larger claims reserving methodology.

\end{abstract}

\section{Introduction}\label{sec:introduction}

With the increasing ease in data collection and greater availability of computational resources for financial institutions, actuarial practitioners and academics face everyday larger quantities of complex data. All this data grants insurers access to detailed information about their insureds and the claims that are typically collected over long periods of time, allowing for a better understanding of their portfolio and, among other, more accurate loss modelling. 

This increase in quantity, however, also leads to an increase in data complexity. Actuaries must sort through large datasets to retrieve relevant information and derive predictions for premiums or the development of claims, using data that is more often than before unstructured, missing or untimely. One particular complexity frequently encountered in insurance is the issue of incomplete or censored data, arising when part of the information is missing for a specific observation. This situation can arise, for example, in a model for coupled lifetimes when death occurs for one of the individuals in the study, as illustrated in \cite{antonio2022}. Claims severities can be censored as well if the insurer benefits from a reinsurance treaty that covers a predefined portion of the losses. We can also consider the case where a given prescription period on the payments made by the insurer has passed, thereby stopping further payments towards the claimant even if the claim is still open. Multiple factors can also cause censoring. In the example with the coupled lifetimes, observations can be censored either by the death of one individual or if they simply revoke their contract with the insurance company. Insurers can thus observe a multitude of censoring scenarios in which one or more variables can be censored or where multiple factors can cause censoring. 

Additionally, a second complexity often encountered in insurance data is the correlation between different - potentially censored - variables of interest to the insurers. This dependence can take the form, for example, of temporal dependence between the occurrence of a specific event and a claim's development delay as in \cite{zhou2010}, \cite{shi2016} and \cite{shi2018}. \cite{frees2008}, \cite{frees2009} or \cite{yang2019} also illustrate hierarchical dependence within multi-features claims. We can even find cross-sectional dependence between losses and expenses, as in \cite{frees1998}, or between a claim's lifetime and its final amount, as in \cite{lopez2019}. 

To account for this, many authors in the actuarial literature have used the framework of copulas, i.e., multivariate distribution functions whose marginals are uniformly distributed on the unit interval. Copulas allow a representation of the joint distribution function of multivariate data by separating their marginal distributions from the inherent dependence structure. Examples of their application abound in the statistical literature, and since their introduction by \cite{sklar1959}, many authors have extensively studied and analyzed their properties. 

Among the different classes of copulas, \cite{schweizer1983} introduced the Archimedean family, widely applied today in many fields beyond insurance. Contrarily to other copula families, Archimedean copulas are not derived from Sklar's theorem, as highlighted in \cite{grosser2021}. They are characterized instead by a function $\phi(.)$, defined with one or more so-called dependence parameters, called the \textit{generator} of the copula. Thus, estimating $\phi(.)$ directly allows us to identify the copula model most appropriate for a given set of multivariate observations. For this purpose, \cite{genest1993} propose a non-parametric approach relying on a study of the bivariate probability integral transformation for Archimedean copulas. Their estimator is, however, only applicable to complete data, seldom encountered in insurance, as mentioned earlier.

Although some authors have worked with \cite{genest1993}'s non-parametric estimator or extensions of it, these contributions are still rare, especially in the actuarial literature. Considering the large amounts of data available to insurers and the highly diverse censoring schemes that these datasets can contain, we aim in this paper to fill this gap by suggesting the use of a non-parametric estimator of Archimedean copula generators applicable to various censoring scenarios. For this purpose, we base ourselves on the approach proposed in \cite{genest1993} and slightly modify it, using results from \cite{wang2000} and an estimator for the join distribution proposed by \cite{akritas2003} in survival analysis. To the best of our knowledge, this estimator has never been used in an actuarial setting nor in the application of a non-parametric estimator for Archimedean copulas.  

The motivation of our work is three-fold. First, we use a non-parametric estimator that readily applies to large amounts of data. The approach we follow in this paper allows insurers to benefit from the large datasets at their disposition and let the data speak for itself to select the most appropriate model. Second, we account for the complexity of this data by using a methodology that encapsulates its multivariate features and any possible censoring schemes, including censoring caused by multiple factors. Third, we gather and provide methods to validate and assess the goodness-of-fit of our estimation, thereby strengthening it.

We structure our paper as follows: first, we provide a brief literature review of the main papers we rely on in our approach in Section~\ref{sec1}. In Section~\ref{sec:Model}, we describe the statistical model for the non-parametric estimator using methods borrowed from survival theory. Section~\ref{sec:validation} presents three model validation approaches under different censoring scenarios. In Section~\ref{sec:simulation}, we test our model within these scenarios using extensive simulation studies. Section~\ref{sec:application} presents an application to a Canadian automobile claims dataset in which policyholders benefit from multiple insurance coverages, and that provides a good illustration of a more complex censoring scheme. To build a micro-level claims reserving model, we use copulas to account for the underlying dependence between coverages. We conclude our work in Section~\ref{sec:conclusion}.

\section{Literature review}\label{sec1}

In recent years, attention has been paid to Archimedean copulas, particularly in finance and insurance. Allowing for very flexible dependence structures, this class of copulas has become popular thanks to their simple form and statistical tractability, among other interesting properties. Notably, many of them allow high-dimensional dependence modeling using only one parameter governing the strength of dependence. Although initially more widely used in life insurance where they arise from frailty models, Archimedean copulas have grown popular in non-life insurance as well. 


Thanks to the large quantities of data more easily available, benefits can be reaped from using a non-parametric estimator such as the one proposed by \cite{genest1993} to directly retrieve the generator of Archimedean copula models. We discuss two interesting examples of the application of this estimator in the case of censored data. First, the work of \cite{denuit2006} combines an estimator for the generator from \cite{wang2000} to an estimator of the joint distribution proposed by \cite{akritas1994} to model the dependence between the allocated loss adjustment expenses and right-censored losses. Their methodology is, however, only applicable to cases where only one variable is subject to censoring, thereby limiting possible replications to more flexible scenarios.


To the best of our knowledge, the only non-parametric approach for more flexible censoring scenarios is the work proposed in \cite{lopez2015}. In their paper, the authors propose a discrete estimator and two smooth estimators for Archimedean copulas and any type of copula applicable to various censoring scenarios. Their discrete estimator extends the empirical copula estimator proposed in \cite{deheuvels1997} by using random weights in the empirical joint distribution to account for the bias caused by censored observations. For the two smooth copula estimators, \cite{lopez2015} introduce kernel functions in addition to the weights of the empirical joint distribution estimator. 

In the present paper, we use an estimator situated between those of \cite{denuit2006} and \cite{lopez2015}. As in \cite{denuit2006}, we focus only on Archimedean copulas and use a non-parametric estimator specifically for Archimedean generators. We, however, seek to allow for the same level of flexibility in the choice of censoring scenarios as \cite{lopez2015}. As such, we extend the work of \cite{denuit2006} to the case where any or all variables can be subject to censoring by replacing the estimator for the joint distribution with that proposed in \cite{akritas2003}.

In addition, we want to use robust approaches to validate the results obtained from such a non-parametric estimator. Goodness-of-fit tests for copulas have been discussed by some authors in the literature, notably \cite{genest2009} who present a review of available tests, with a focus on so-called \textit{blanket} tests, i.e., that are not constrained to one specific class of copulas or that do not require excessive data manipulation, based on the empirical copula and Kendall distribution. The authors also propose a new testing procedure based on Rosenblatt's transform. More recently, \cite{grosser2021} reviewed all the latest developments in copula theory, including estimation and testing methods. However, all these tests only apply to complete data. The literature on goodness-of-fit tests for copulas with censored data is much scarcer. For the specific case of Archimedean copulas, \cite{lakhal2010} proposes a test based on a Cramer-von-Mises type distance and Kendall distribution applicable to Archimedean copulas. \cite{wang2010} proposes a test based on a multiple-imputation procedure of the censored data for that same class of copulas, and \cite{emura2010} use a cross-ratio function. Moving away from Archimedean copulas, \cite{yilmaz2011} derive likelihood and pseudo-likelihood tests, embedding the copulas under the null and alternative hypotheses. \cite{lin2020} propose a general smooth test that assumes a specific form for the alternative copulas. Others, such as \cite{chen2008}, describe a goodness-of-fit test easily applicable to high-dimensional copulas, and that allows for any parametric copula specification. They propose a penalized pseudo-likelihood ratio statistic to compare various semi-parametric multivariate survival functions subject to copula misspecification and general censorship. Recently, \cite{zhou2021} and \cite{sun2023} presented general tests based on information matrices applicable to various copula families. While \cite{zhou2021} remains focused on right-censored data, \cite{sun2023} allow for greater flexibility by including interval-censoring and recurrent events in their test. Finally, in addition to their three non-parametric copula estimators, \cite{lopez2015} also propose a test similar to that of \cite{lakhal2010} that can be used with different kinds of distances such as Kolmogorov-Smirnov, Cramer-von-Mises or the square-root of a quadratic integrated distance. 

In this paper, we use three approaches to validate the non-parametric estimation of the Archimedean copula generator. More specifically, we use an omnibus procedure based on the likelihood function, a method based on the $L^2$-norm distance, inspired by \cite{lakhal2010} and \cite{lopez2015}, and the test proposed by \cite{wang2010}.

\section{Statistical Model}
\label{sec:Model}
In this section, we describe an approach to retrieve the Archimedean copula generator $\phi(.)$ in the case of flexible censoring. More specifically, we present a non-parametric estimator building on the work of \cite{akritas2003} in survival analysis. The advantage of using such a non-parametric estimator lies in the possibility of directly tailoring the model to the data, even in the presence of complex censoring. This contrasts with the more naive copula modelling approach that usually consists of fitting a certain number of copulas and comparing their performance using information criteria such as the Akaike Information Criterion (AIC) or Bayesian Information Criterion (BIC). 

We begin this section by defining the notation before describing the estimator. We end by detailing the graphical comparison method, thanks to which we identify the copula model most appropriate to a given dataset. For clarity, we focus solely on the bivariate case, but the notation and model described hereafter can easily be extended to a multivariate framework.

\subsection{Notation}
\label{sec:notation}
We define the following model components:
\begin{itemize}
    \item $\vect{T}=(T_1,T_2)$ with $T_i \in \mathbb{R}^+$ for $i=1,2$ is the initial vector of variables of interest. They can represent times-to-event, losses, etc. To avoid identifiability issues with the copula, we assume these variables to be continuous; 
    \item $\vect{X}=(X_1,X_2)$ with $X_i \in \mathbb{R}^+$ for $i=1,2$ is the vector of censoring variables, such that $\vect{T}$ and $\vect{X}$ are independent;
    \item $\vect{\omega}=(\omega_1, \omega_2)$, with $\omega_i > 0$ for $i=1,2$ are constants acting as additional censoring factors on $\vect{T}$. We call these the \textit{limits}.
    \item $\vect{Y} = (Y_1, Y_2)$ where $Y_i = \min(T_i, X_i, \omega_i)$ for $i=1,2$ is the observed bivariate vector, taking censoring of $\vect{T}$ into account;
    \item $\vect{\Delta} = (\mathds{1}_{[Y_1 = T_1]}, \mathds{1}_{[Y_2 = T_2]})$ is the vector of censoring indicators.
\end{itemize}

To illustrate the difference between the censoring variable $X$ and the limit $\omega$, consider a study of the time to recovery after administration of a treatment of patients suffering from a medical condition. The time to recovery, denoted by $T$, is censored if the patient dies, with the time to death being the censoring variable $X$. In addition, health professionals might consider the treatment ineffective if the patient does not recover after a given amount of time, even if he is still alive. Further treatments will then need to be considered. In this example, the time elapsed until the treatment is deemed to have failed is the limit $\omega$. Even if the patient is still alive, i.e., $T > X$, the imposition of $\omega$ will censor his time to recovery. The observed delay $Y=\min(T, X, \omega)$ is of interest to medical practitioners. $Y$ represents the delay until the first event, being either full recovery, death, or need for additional treatments. 

Further applications can be found in finance, insurance, and other fields. Section~\ref{sec:application} focuses on an application to bivariate claims delays. 

\subsection{Non-parametric estimator of the generator}
\label{sec:estimator}
We now describe the non-parametric approach leading to the identification of the copula linking the pairs of censored variables $(Y_1, Y_2)$.

Assuming flexible censoring scenarios where either one or both variables can be subject to random censoring, we follow the methodology laid out in \cite{akritas2003} and work with the following extension of \cite{beran1981}'s estimators for the conditional distributions of $Y_1$ (resp. $Y_2$) given $Y_2=y_2$ (resp. $Y_1 = y_1$):
\begin{align}
\label{eq:beran}
    \hat{F}_{1 \vert 2}(y_1  \vert  y_2) = 1 - \prod_{Y_{i1} \leq y_{1}, \Delta_{i1} = 1} \Bigg( 1 - \frac{W_{ni2}(y_{2};h_{n})}{\sum_{j=1}^{n}W_{nj2}(y_{2};h_{n})\mathds{1}_{Y_{j1} \geq Y_{i1}}}\Bigg),
\end{align}
where $y_2$ must be uncensored and
\begin{align*}
	W_{ni}(y;h_{n}) = \begin{cases}  \frac{k\Big(\frac{y-Y_{i2}}{h_{n}}\Big)}{\sum_{\Delta_{j1}=1}k\Big(\frac{y-Y_{j2}}{h_{n}}\Big)},&\hspace{0.3cm}\text{if}\hspace{0.2cm} \Delta_{i2}=1 \\ 0,&\hspace{0.3cm}\text{if}\hspace{0.2cm} \Delta_{i2}=0.  \end{cases}
\end{align*}
Here, $k(.)$ is a known kernel function and $\{h_n\}$ is the bandwidth: a sequence of positive constants such that $h_n \rightarrow 0$ as $n \rightarrow \infty$. We work with a similar estimator for $\prob[Y_2 \leq y_2  \vert  Y_1 = y_1] = \hat{F}_{2 \vert 1}(y_2  \vert  y_1)$.

The estimator of the joint distribution proposed in this context by \cite{akritas2003} is then given by:
\begin{align}
\label{eq:joint}
    \hat{F}(\vect{y}) = w(\vect{y})\int_0^{y_2} \hat{F}_{1 \vert 2}(y_1  \vert  z_2)d\Tilde{F}_2(z_2) + (1 - w(\vect{y}))\int_0^{y_1} \hat{F}_{2 \vert 1}(y_2  \vert  z_1)d\Tilde{F}_1(z_1),
\end{align}
where $\Tilde{F}_{1}$ and $\Tilde{F}_{2}$ are the marginal estimators of Kaplan and Meier (1958): 
\begin{align*}
	\Tilde{F}_{j}(y_{i}) = 1 - \prod_{Y_{i,j} \leq y_{i}, \Delta_{ij}=1} \Big(1 - \frac{1}{n-i+1}\Big), \hspace{0.3cm}j=1,2.
\end{align*}
    The weights $w(\vect{y})$ minimise the mean-squared error of $\hat{F}(\vect{y})$. Equation~\eqref{eq:joint} is the estimator that, combined with \cite{genest1993}'s approach, allows us to retrieve the generator of the copulas in the presence of different censoring scenarios. We start from \cite{genest1993}' proposition laid out just below. \\~\\
\textbf{Proposition 1.1.} [\cite{genest1993}] \textit{Let $S_1(t_1)$ and $S_2(t_2)$ be marginal survival functions whose dependence function $C(.)$ is of the form
\begin{align*}
    C(S_1(t_1),S_2(t_2)) = \phi^{-1}\Big\{\phi(S_1(t_1)) + \phi(S_2(t_2))\Big\},
\end{align*}
where $\phi(.) \in (0,1]$ is a convex decreasing function such that $\phi(1)=0$. Let
\begin{align*}
    U = \frac{\phi(S_1(t_1))}{\phi(S_1(t_1)) + \phi(S_2(t_2))} \hspace{0.3cm}\text{and}\hspace{0.3cm} V = \phi^{-1}\Big\{ \phi(S_1(t_1)) + \phi(S_2(t_2)) \Big\} = C(S_1(t_1),S_2(t_2)).
\end{align*}
Then, 
\begin{enumerate}
    \item $U$ is uniformly distributed on $(0,1)$;
    \item $V$ follows the Kendall distribution defined as $K(\nu) = \nu - \frac{\phi(\nu)}{\phi^{(1)}(\nu)}$, for $0<\nu \leq 1$; and
    \item $U$ and $V$ are independent. 
\end{enumerate}}

\cite{genest1993} suggest the following estimator for the Kendall distribution in the presence of complete data:
\begin{align*}
    \hat{K}_n(\nu) = \frac{1}{n}\# \{i \vert \nu_i \leq \nu\},
\end{align*}
where $\#$ denotes the cardinality of a set.

We cannot use this estimator in the presence of censored data. Instead, we choose to apply the extension proposed by \cite{wang2000} to estimate the Kendall distribution by
\begin{align}
\label{eq:kendall}
    \hat{K}_n(\nu) = \int_0^\infty \int_0^\infty \mathds{1}_{[\hat{F}(\vect{y}) \leq \nu]}d\hat{F}(\vect{y}),
\end{align}
in which we insert \eqref{eq:joint} as the estimator of the joint distribution, allowing for all possible censoring scenarios.

Finally, we insert $\hat{K}_n(\nu)$ in the estimator for the Archimedean generator proposed by \cite{genest1993}:

\begin{align}
\label{eq:2}
    \hat{\phi}_n(\nu) = \exp \Bigg\{\int_{\nu_0}^\nu \frac{1}{t-\hat{K}_n(t)}dt \Bigg\},
\end{align}
with $0<\nu_0<1$ an arbitrarily chosen constant. Thanks to the use of Equation~\eqref{eq:joint}, we can retrieve the generator non-parametrically, regardless of the level of censoring present in the data.

\subsection{Graphical comparison}
Starting from the empirical estimator $\hat{K}_n(\nu)$, we illustrate the graphical procedure introduced by \cite{genest1993} to select the best fitting model among competing copulas for a given dataset. We define the univariate function $\lambda(\nu)$ from which \cite{genest1993} recover the generator in Equation~\eqref{eq:2} as
\begin{align*}
    \lambda(\nu) = \frac{\phi(\nu)}{\phi^{(1)}(\nu)}.
\end{align*}
An estimator of $\lambda(\nu)$ is easily retrieved by 
\begin{align}
\label{eq:1}
    \hat{\lambda}_n(z) = z - \hat{K}_n(z).
\end{align}
The idea is now to graphically compare the empirical estimator $\hat{\lambda}_n(\nu)$ to the equivalent functions $\lambda_{\hat{\alpha}}(\nu)$ of each competing copula model under consideration, where $\hat{\alpha}$ is the dependence parameter estimated with \eqref{eq:3}. Table~\ref{tab:lambda} presents some of the most popular Archimedean copulas and the one-to-one relationship between their dependence parameter and Kendall's tau. The estimation of $\hat{\tau}$ is straightforward:
\begin{align}
\label{eq:3}
    \hat{\tau} = 4 \int_0^1 \hat{\lambda}_n(\nu)d\nu = 3 - 4\int_0^1 \hat{K}_n(\nu)d\nu.
\end{align}
\begin{table}[h]
\begin{center}
\begin{minipage}{\textwidth}
\caption{$\lambda_\alpha(\nu)$ for some popular Archimedean copulas.}
\label{tab:lambda}
\scalebox{0.87}{
\begin{tabular}{@{}l c c @{}}
\toprule
Copula & $\tau$ &$\lambda_\alpha(\nu$) \\
\midrule
Clayton & $\alpha/(\alpha +2)$ &$\nu(\nu^\alpha-1)/\alpha$ \\
Frank & $1 + \frac{4}{\alpha}\Big( \int_0^\alpha \frac{\xi}{\alpha(e^{\xi}-1)}d\xi - 1 \Big)$ &$-\ln{\Big\{(e^{-\alpha \nu}-1)/(e^{-\alpha}-1)}\Big\} / (-\alpha / (e^{\alpha \nu}-1))$ \\
Gumbel-Hougaard & $1 - 1/\alpha$ &$\nu \ln({\nu})/\alpha$ \\
Joe & - &$-\ln{\Big\{1-(1-\nu)^\alpha}\Big\} / - \Big((\alpha (1-\nu)^{(\alpha-1)})/(1-(1-\nu)^\alpha)\Big)$ \\
\bottomrule
\end{tabular}}
\end{minipage}
\end{center}
\end{table}

To illustrate the methodology laid out in this section, we consider the Loss-ALAE dataset used in \cite{frees1998}, \cite{denuit2006}, and \cite{lopez2015}. We are in a single-censoring scenario where only the loss variable is subject to right-censoring. We estimate $\hat{K}_n(\nu)$ and $\hat{\lambda}_n(\nu)$ using, respectively, Equation~\eqref{eq:kendall} and Equation~\eqref{eq:1}. We then plot $\hat{\lambda}_n(\nu)$ and $\lambda_{\hat{\alpha}}(\nu)$ for the Clayton, Frank, Gumbel, and Joe copulas. We estimate $\hat{\alpha}$ for each using the one-to-one relationship between Kendall's tau, retrieved from Equation~\eqref{eq:3} using $\hat{\lambda}_n(\nu)$ and the dependence parameters, as illustrated in Table~\ref{tab:lambda}. 

\begin{figure}[h!]
\centering
\includegraphics[width=.65\textwidth]{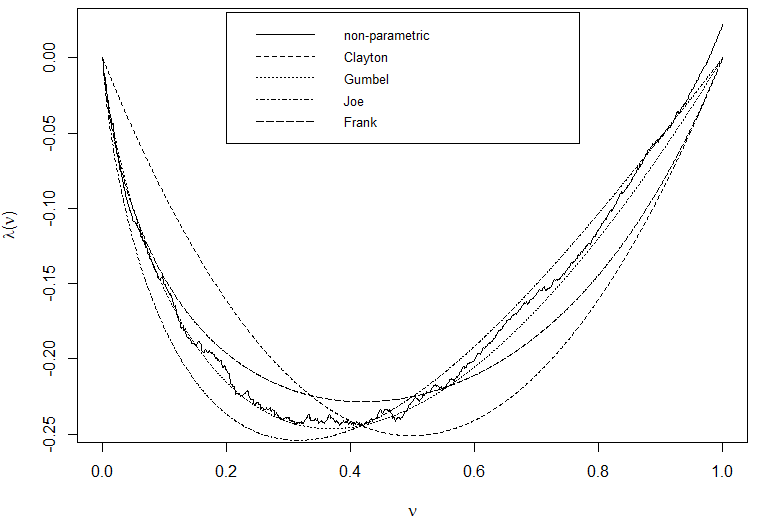}
\caption{Graphical comparison of different copulas for the Loss-ALAE dataset.}
\label{fig:denuit}
\end{figure}

Figure~\ref{fig:denuit} compares the empirical estimate and the copulas under consideration. We observe that when using the approach laid out in Section~\ref{sec:estimator} with \cite{akritas2003}'s estimator in \eqref{eq:joint} for the joint distributions, we obtain very similar results to those presented in \cite{denuit2006} who used the estimator from \cite{akritas1994}. More specifically, we clearly see in Figure~\ref{fig:denuit} that $\hat{\lambda}_n(\nu)$ closely follows the curve $\lambda_{\hat{\alpha}}(\nu)$ of the Gumbel copula. This indicates that the Gumbel copula best fits the Loss-ALAE data. We also show in Table~\ref{tab:comp} the estimated Kendall's tau and dependence parameters obtained in, respectively, \cite{denuit2006} using \cite{akritas1994}'s estimator in the second column (single censoring model), compared to those in the third column obtained using the methodology from Section~\ref{sec:estimator} (flexible censoring model). We observe that all estimated parameters are very similar. This illustrates the adequacy of the approach that we use in this paper.

\begin{table}[h]
\begin{center}
\caption{Comparison of parameters estimates.}
\label{tab:comp}
\begin{tabular}{@{}l c c @{}}
\toprule
 & Single censoring model & Flexible censoring model\\
Copula & $\tilde{\tau}=0.3567$ & $\hat{\tau}=0.3507$ \\
\midrule
Clayton & 1.1088 & 1.0803 \\
Frank & 3.5919 & 3.5177\\
Gumbel-Hougaard & 1.5544 & 1.5403 \\
Joe & 2.0074 & 1.9805 \\
\bottomrule
\end{tabular}
\end{center}
\end{table}

\section{Results validation}
\label{sec:validation}
While practical and easy to implement, the estimator described in Section~\ref{sec:estimator} is non-parametric, and the choice of the copula model is driven at this stage by a graphical analysis only, as illustrated in Figure~\ref{fig:denuit}. Goodness-of-fit tests for copulas are rare in the literature and even more so in the presence of censoring. To validate our choice and increase the level of confidence in our approach, we gather three different approaches in this section: a pseudo-maximum likelihood estimation procedure of the dependence parameters, a validation based on the $L^2$-norm distance, and a statistical test for Archimedean copulas with censored data introduced by \cite{wang2010}.

\subsection{Omnibus estimation procedure}
\label{sec:omnibus}
The omnibus estimation procedure, or maximum pseudo-likelihood procedure, is a semi-parametric optimization procedure that substitutes empirical versions of the marginal distributions in the (parametric) likelihood function of the copula model, $L(.)$. 

We use rescaled Kaplan-Meier estimators for $T_1$ and $T_2$, multiplying the original Kaplan-Meier versions by $n/(n+1)$, where $n$ is the total number of observations in the sample. As mentioned in \cite{genest1995} as well as \cite{denuit2006}, this rescaling prevents issues related to the potential unboundedness of the copula log-density as $u_1$ and $u_2$ tend to one. 

Once we have the marginal distribution functions, we use them in the estimation step, i.e., the optimization of the model likelihood to obtain the pseudo-likelihood estimator $\hat{\alpha}^*$ of the dependence parameter:
\begin{align*}
    \hat{\alpha}^* = \argmax L(u_1, u_2, &\delta_1, \delta_2; \alpha),
\end{align*}
where the likelihood function under flexible censoring scenarios is given by:
\begin{equation} \label{eq:lik}
  \begin{split}
     L(u_1, u_2; &\delta_1, \delta_2; \alpha) = \prod_{i=1}^n c_\alpha(u_{1i},u_{2i};\alpha)^{\delta_{1i}\delta_{2i}} + \bigg(\frac{\partial C_\alpha(u_{1i},u_{2i};\alpha)}{\partial u_1}\bigg)^{\delta_{1i}(1-\delta_{2i})} \\&+ \bigg(\frac{\partial C_\alpha(u_{1i},u_{2i};\alpha)}{\partial u_2}\bigg)^{(1-\delta_{1i})\delta_{2i}} + C_\alpha(u_{1i},u_{2i};\alpha)^{(1-\delta_{1i})(1-\delta_{2i})}.
  \end{split}
\end{equation}

$C_\alpha$ is the candidate Archimedean copula under consideration, $c_\alpha$ its density and $\delta_{ji} = 1$ if $T_j < X_j$ for $j=1,2$ for observation $i$ and equals zero otherwise (i.e., for a censored observation). The partial derivatives with respect to either $u_1$ or $u_2$ can be found in Table~\ref{tab:derivatives}. 

\begin{table}[h]
\begin{center}
\caption{Partial derivatives for some popular Archimedean copulas with $\tilde{u} = -\ln u$ and $\bar{u}=1-u$.}
\label{tab:derivatives}
\begin{tabular}{@{}l c  @{}}
\toprule
Copula & $\partial C_\alpha(u_1.u_2)/\partial u_1$ \\
\midrule
Clayton & $[1+u_1^\alpha(u_2^{-\alpha}-1)]^{-1-1/\alpha}$ \\
Frank & $[e^{-\alpha u_1} - e^{-\alpha(u_1+u_2)}] \times [(1-e^{-\alpha}) - (1-e^{-\alpha u_2})(1-e^{-\alpha u_1})]^{-1}$ \\
Gumbel & $u_1^{-1}e^{\{-(\Tilde{u}_2^\alpha+\Tilde{u}_1^\alpha)^{1/\alpha}\}}[1+(\Tilde{u}_2/\Tilde{u}_1)^\alpha]^{-1+1/\alpha}$ \\
Joe & $(1-\Bar{u}_2^\alpha)(1-\Bar{u}_2^\alpha + \Bar{u}_2^\alpha \Bar{u}_1^{-\alpha})^{-1+1/\alpha}$ \\
\bottomrule
\end{tabular}
\end{center}
\end{table}

In order to use the omnibus procedure to validate the model choice resulting from the graphical comparison, we retrieve the pseudo-maximum likelihood estimator $\hat{\alpha}^*$ for the chosen candidate copula models and compare them to the estimators $\hat{\alpha}$ found with the non-parametric model using \eqref{eq:3} and Table~\ref{tab:lambda}. The copula with the smallest difference between these two estimators is the most appropriate for the data.

\subsection{$L^2$-norm}
\label{sec:L2norm}
The second approach to validate the choice of copula model is based on the $L^2$-norm. We follow the methodology laid out in \cite{wang2000} that introduce a goodness-of-fit statistic specifically designed for Archimedean copulas with censored data under the assumption that a consistent non-parametric estimator of the bivariate joint distribution function is available. Note that with estimated parameters present in the specification of the null hypothesis for this test, the approach we show here is not an actual statistical test but rather a way to validate our model. However, we will use the word \textit{test} in this section for simplicity. As mentioned in the introduction, more formal and general methods have been proposed in the literature. We discuss one in Section~\ref{sec:wang}.  

\cite{wang2000}'s test relies on the $L^2$-norm distance between the empirical estimator $\hat{K}_n(\nu)$ and the corresponding $K_{\hat{\alpha}}(\nu)$ of the hypothesized parametric copula model,
\begin{align*}
    D(\hat{\alpha}) = \int_\xi^1 \big( \hat{K}_n(\nu) - K_{\hat{\alpha}}(\nu)\big)^2 d\nu.
\end{align*}
We use Riemann sum approximations to simplify numerical analysis: 
\begin{align*}
    D(\hat{\alpha}) = \sum_{i=1}^{n} \big( \hat{K}_n(\nu_{(i)}) - K_{\hat{\alpha}}(\nu_{(i)})\big)^2(\nu_{(i)} - \nu_{(i-1)}),
\end{align*}
where $\nu_{(i)}$ is the $i^{\text{th}}$ ordered value of $\{\nu_1,\nu_2,...,\nu_n \}$. However, the asymptotic distribution of $D(\hat{\alpha})$ is difficult to retrieve analytically. We propose below a parametric bootstrap procedure inspired by \cite{lakhal2010} and \cite{lopez2015} to select the most appropriate copula model based on a \textit{pseudo} $p$-value:\\~\\
\textit{Step 1.} Compute $\hat{K}_n(\nu)$ for the data at hand and, for the $M$ candidate models under consideration, estimate $\hat{\alpha}_m$, for $m=1,...,M$. \\
\textit{Step 2.} For each candidate model $C_{\hat{\alpha}_m}$, generate $B$ censored samples of size $n$.\\
\textit{Step 3.} For each of the $B$ samples, estimate $\hat{\alpha}_{m,b}$ and $D(\hat{\alpha}_{m,b})$. \\
\textit{Step 4.} For each model $m$, obtain the pseudo $p$-value as 
\begin{align*}
    p_m = \frac{1}{B}\sum_{b=1}^{B}\mathds{1}[\min_l D(\hat{\alpha}_{b,l}) > D(\hat{\alpha}_{b,m})]
\end{align*}
with $m \neq l$. The copula model $m$ best fitted to the data is the one for which the pseudo $p$-value is the smallest. In this case, we can interpret $p_m$ as the number of times among $B$ replications that copula model $m$ is selected as most fitted for the data at hand.

\subsection{Goodness-of-fit test for censored dependent data}
\label{sec:wang}
The omnibus procedure and $L^2$-norm validation approach from Sections \ref{sec:omnibus} and \ref{sec:L2norm} are easy to implement and give a good idea of whether a copula model is appropriate for the data at hand. However, neither are formal statistical tests. To further increase the confidence in our choice of copula model, we present a more formal approach for Archimedean copulas, proposed by \cite{wang2010}. 

This goodness-of-fit test applies to uncensored and right-censored data thanks to a multiple imputation procedure. It bases its premise on Proposition 1.1. from \cite{genest1993}.

For a bivariate vector of dependent observations $(T_1,T_2)$ whose marginal survival functions $S_1(.)$ and $S_2(.)$ have a dependence function $C(.)$ of the form
\begin{align*}
    C(S_1(t_1),S_2(t_2)) = \phi^{-1}\Big\{\phi(S_1(t_1)) + \phi(S_2(t_2))\Big\},
\end{align*}
then, the random variables
\begin{align}
\label{eq:UV}
    U = \frac{\phi(S_1(t_1))}{\phi(S_1(t_1)) + \phi(S_2(t_2))} \hspace{0.3cm}\text{and}\hspace{0.3cm} V = \phi^{-1}\Big\{ \phi(S_1(t_1)) + \phi(S_2(t_2)) \Big\} = C(S_1(t_1),S_2(t_2)).
\end{align}
are independent. In other words, under the null hypothesis that an Archimedean copula can model the dependence between $(T_1, T_2)$ with generator $\phi(.)$, the correlation coefficient between $U$ and $V$ is null: 
\begin{align*}
    H_0: \rho = 0 \hspace{0.3cm} \text{vs} \hspace{0.3cm} H_1: \rho \neq 0.
\end{align*}

Let 
\begin{align*}
    r_n = \frac{\sum_{i=1}^{n} (\hat{U}_i - \Bar{\hat{U}})(\hat{V}_i - \Bar{\hat{V}})}{\sqrt{\sum_{i=1}^{n}(\hat{U}_i - \Bar{\hat{U}})^2 \sum_{i=1}^{n}(\hat{V}_i - \Bar{\hat{V}})^2}}
\end{align*}
where $\hat{U}$ and $\hat{V}$ are consistent estimators of $U$ and $V$ and where $\Bar{\hat{U}}$ and $\Bar{\hat{V}}$ are the sample means of $\hat{U}_i$ and $\hat{V}_i$. The test statistic is defined as 
\begin{align*}
    Z_n = \frac{1}{2}\log \Big[ \frac{1 + r_n}{1-r_n} \Big]
\end{align*}
and we have that $\sqrt{n}Z_n \rightarrow N(0,1)$ in distribution. 

For censored data, we can not directly retrieve $\hat{U}_i$ and $\hat{V}_i$ from the estimators proposed by \cite{genest1993} based on Equation~\eqref{eq:UV}. Instead, \cite{wang2010} shows that, depending on the censoring pattern, the data can be simulated using a multiple imputation procedure from one of the distributions presented in \ref{app:1}. All proofs and further details are in \cite{wang2010}. 


\section{Simulation study}
\label{sec:simulation}
Using simulation studies, we now illustrate the approach described in Section~\ref{sec:Model}. We provide in this section a few examples to showcase the graphical model selection procedure and the validation approaches introduced in Section~\ref{sec:validation}. 

\subsection{Independence case}
The first way to validate the results from the estimator for the generator in Section~\ref{sec:Model} is to analyze how it handles independent data. Setting Kendall's tau equal to 0 in Table~\ref{tab:lambda}, each of the four copulas should converge to the independence copula. This is the case if $\alpha = 0$ for both the Clayton and Frank copulas and $\alpha = 1$ for the Gumbel and Joe copulas. 

\begin{figure}[h!]
\centering
\includegraphics[width=.6\textwidth]{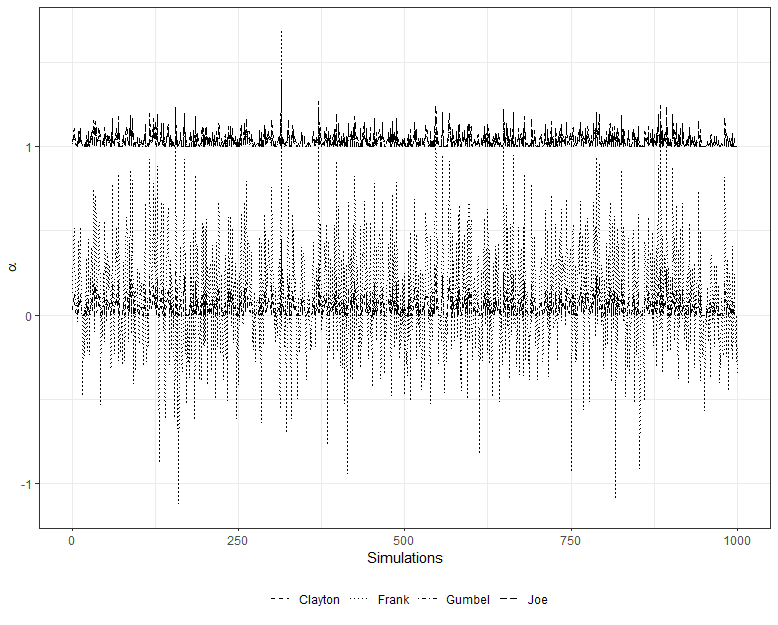}
\caption{Estimated dependence parameters for different copulas in case of independence ($\tau = 0$).}
\label{fig:tau_ind}
\end{figure}

We simulate samples of size $n=1000$ from each of the four copulas with $\tau = 0$. We then use the methodology laid out in Section~\ref{sec:Model} to estimate $\hat{K}_n(\nu)$, $\hat{\tau}$ and, finally, $\hat{\alpha}$ for each copula. Table~\ref{fig:tau_ind} presents the results of 1000 simulations where we plot the estimated dependence parameters. We also present the average estimates over the 1000 simulations in Table~\ref{tab:alpha}.

Figure~\ref{fig:tau_ind} shows that the Clayton and Frank copulas simulations are clearly centered around $0$, with the Frank copula presenting slightly larger variations. These are, however, simply due to the wider range of possible values that $\alpha$ can take for this specific copula. The simulations are very close to 1 for the Gumbel and Joe copulas. Table~\ref{tab:alpha} confirms these results. Our approach using \cite{akritas2003}'s estimator for the joint distribution provides adequate estimates of the dependence parameters when $\tau = 0$. 

\begin{table}[h]
\begin{center}
\caption{Average estimates of $\hat{\alpha}$ for $1000$ simulations when $\tau = 0$.}
\label{tab:alpha}
\begin{tabular}{@{}c c c c c@{}}
\toprule
 & Clayton & Frank & Gumbel & Joe \\
\midrule
$\hat{\alpha}$ & 0.0249 & 0.0352 & 1.0163 & 1.0150 \\
\bottomrule
\end{tabular}
\end{center}
\end{table}

\subsection{Graphical comparison}
We now move away from the independence scenario and focus on the non-parametric estimator of $\hat{\lambda}_n(\nu)$ used for the graphical comparison. 

We simulate bivariate samples $(T_1, T_2)$ of $n=1000$ observations from the Clayton, Gumbel, Frank, and Joe copulas. In each case, we select the dependence parameter such that Kendall's tau equals $0.4$, and we work with unit-exponential distributions for the marginals. We simulate the vector of censoring variables $(X_1, X_2)$ from exponential distributions with their parameters set such that $20\%$ of observations have at least one censored component. We are in a double-censoring scenario where one or both variables can be censored.

\begin{figure}[h!]
\centering
\includegraphics[width=1\textwidth]{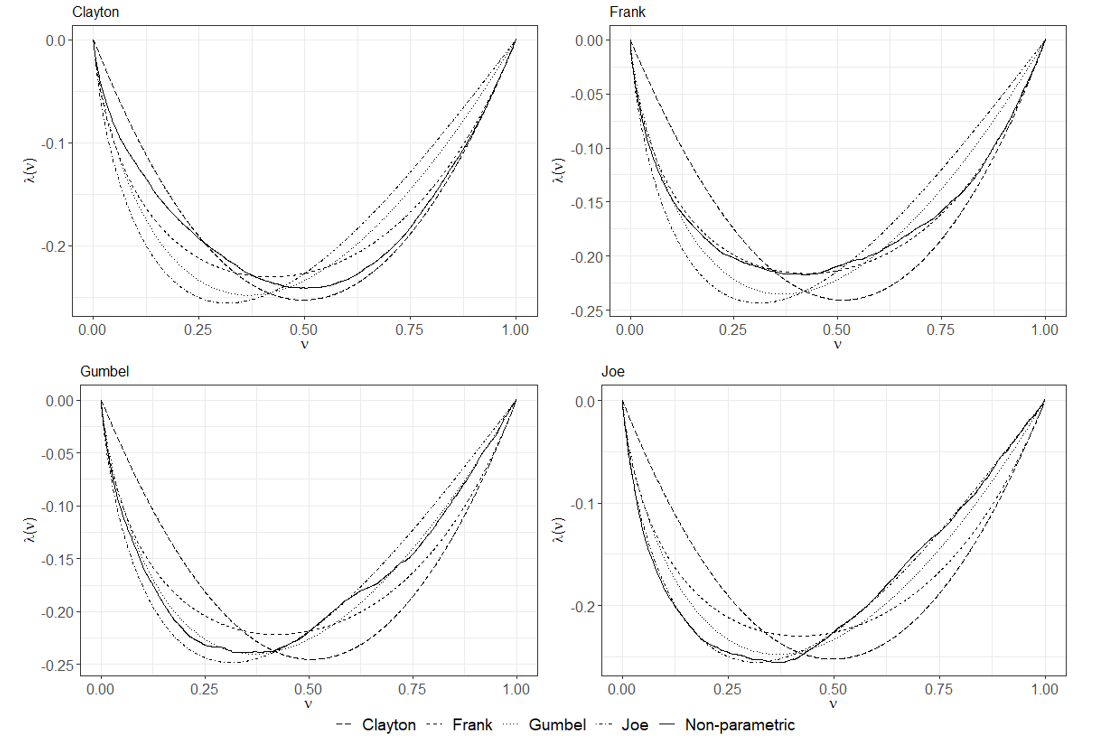}
\caption{Graphical comparison of the estimated $\hat{\lambda}(\nu)$ functions for the four simulated samples. We simulate $n=1000$ bivariate observations from a Clayton (top left), Frank (top right), Gumbel (bottom left) and Joe (bottom right) copulas. }
\label{fig:SIM1}
\end{figure}

Figure~\ref{fig:SIM1} displays the empirical estimator $\hat{\lambda}_n(\nu)$ and the curves of $\lambda_{\hat{\alpha}}(\nu)$ for the four copulas under consideration, for each of the four simulated samples. In each plot, the continuous curve depicts the non-parametric estimator presented in Equation~\eqref{eq:1}. For all, $\hat{\lambda}_n(\nu)$ appears to be closest to the curve of the correct copula, i.e., the copula from which the sample was simulated. In addition to the one from Figure~\ref{fig:denuit}, these graphical comparisons suggest that the approach proposed in Section~\ref{sec:Model} performs quite well under flexible censoring scenarios. In the remainder of this section, we propose further validating these results using three different approaches. 

\subsection{Omnibus procedure}
We begin validating the results from Figure~\ref{fig:SIM1} with the omnibus procedure. In addition to the double-censoring scenario considered here, we add two other censoring schemes: one with complete data (no censoring) and one where we allow for only one censored variable. We work again for these two additional scenarios with simulated samples of size $n=1000$, Kendall's tau equal to $0.4$, and unit-exponential margins. For the single-censoring scenario, we simulate the censoring variable following an exponential distribution such that around $20\%$ of observations have a censored component, similar to the double-censoring scheme.

For each of the three scenarios, Table~\ref{tab:sim1_omni} displays the results of the omnibus procedure. For each simulated sample, we compare the dependence parameters $\hat{\alpha}$ retrieved from the estimation of Kendall's tau in Equation~\eqref{eq:3} to the maximum likelihood estimator $\hat{\alpha}^*$. We observe that in most cases, the distance between $\hat{\alpha}$ and $\hat{\alpha}^*$ is minimal for the correct copula, i.e., the omnibus procedure leads to selecting the copula model from which the data was originally simulated. We count three occurrences where the procedure selects an alternative model. In the single-censoring scenario, the Gumbel model is chosen for the Clayton sample, and the Joe copula is selected for the Gumbel sample. The Gumbel copula is selected for the Frank sample in the double-censoring scenario. However, we observe that the correct copulas present in each case are the second smallest distance between $\hat{\alpha}$ and $\hat{\alpha}^*$ and that the difference with the third and fourth distances is quite marked. For the Gumbel sample in the single censoring scenario for example, the smallest distance between the parameters is equal to $0.0428$ for the Joe copula, $0.0992$ for the Gumbel copula and then, $0.3988$ and $0.4971$ for, respectively, the Clayton and Frank copulas. 

To increase the confidence in model selection, we show in Table~\ref{tab:OMNISIM} the results of performing the omnibus comparison between $\hat{\alpha}$ and $\hat{\alpha}^*$ not once but one thousand times. Each value in this table represents the percentage of simulations in which the competing models were not considered best-fitted for the sample in terms of the difference between the estimated dependence parameters. We observe, for example, that when we simulate a sample from the Clayton copula with $\tau = 0.4$ in a double-censoring scenario, only $276$ simulations out of a thousand reject the Clayton model as best-fitted. Among these $276$ simulations, $188$ opt for a Joe model, $71$ for a Gumbel model, and $17$ for a Frank model. 

The results displayed in Tables \ref{tab:sim1_omni} and \ref{tab:OMNISIM} illustrate that the omnibus procedure is a good first indicator of the appropriateness of the copula model selected via the graphical procedure.

\begin{table}[h]
\begin{center}
\begin{minipage}{540pt}
\caption{Omnibus procedure for different censoring scenarios.}
\label{tab:sim1_omni}
\scalebox{0.9}{
\begin{tabular}{@{}l l c c c c c c c c c @{}}
\toprule
\multirow{2}{*}{True copula} & \multirow{2}{*}{Candidate model} & &  \multicolumn{2}{c}{No censoring} & & \multicolumn{2}{c}{single-censoring}&& \multicolumn{2}{c}{double-censoring} \\ \cline{4-5} \cline{7-8} \cline{10-11}
& & & $\hat{\alpha}$ & $\hat{\alpha}^*$ & & $\hat{\alpha}$ & $\hat{\alpha}^*$ & & $\hat{\alpha}$ & $\hat{\alpha}^*$ \\
\midrule
\multirow{4}{*}{Clayton ($\alpha = 1.3332$)}& Clayton & & \textbf{1.7694} & \textbf{1.8284} & & 1.1789 & 1.3263 & &  \textbf{0.9240} & \textbf{0.8541}\\
&Frank  &  & 5.2063 & 5.7575 & & 3.7724 & 4.1154 & & 2.1323 & 2.7655\\
&Gumbel&  & 1.8846 & 1.7294 & & \textbf{1.5895} & \textbf{1.4926}&  &1.4621 & 1.3070\\
&Joe  &  & 2.6347 & 1.7373 & & 2.0732 & 1.4713 & &1.5231 & 1.2792\\
\midrule
\multirow{4}{*}{Frank ($\alpha = 4.1611$)} & Clayton & & 1.3202 &  0.8049 & & 1.2055 & 0.7428 & &  1.0034 &  0.5326\\
& Frank  &  & \textbf{4.1284} & \textbf{4.2527} & & \textbf{3.8403} & \textbf{3.9827} &  &3.3149 & 2.9126\\
& Gumbel& &  1.6602 & 1.5274 & & 1.6029 & 1.4533 &  &\textbf{1.5018} & \textbf{1.1694}\\
& Joe  &  & 2.2066 & 1.6870 & & 2.0983 & 1.6260 & & 1.9084 & 1.1759\\
\midrule
\multirow{4}{*}{Gumbel ($\alpha = 1.6667$)} & Clayton & & 1.1377 &  0.6557 & & 1.1380 & 0.7392 & &  0.9464 &  0.6493\\
& Frank  & & 4.1641 & 4.3241 & & 3.6674 & 4.1645 &  &2.8121 & 3.6805\\
& Gumbel&  & \textbf{1.5690} & \textbf{1.5934} & & 1.5691 & 1.6683 &  &\textbf{1.4734} & \textbf{1.5283}\\
& Joe  &  & 2.0345 & 1.8735 & & \textbf{2.0347} & \textbf{1.9919} & & 1.8671 & 1.7554\\
\midrule
\multirow{4}{*}{Joe ($\alpha = 2.2191$)} & Clayton & & 1.2609 & 0.6067 & & 1.1494 & 0.5514 & &  0.9344 & 0.5170\\
& Frank  &  & 3.9802 & 4.2203 & & 3.6968 & 3.8131 &  &3.1298 & 3.4508\\
& Gumbel & & 1.6305 &  1.7284 & & 1.5748 & 1.6356 &  &1.4674 & 1.5638\\
& Joe   & & \textbf{2.151} & \textbf{2.1997} & & \textbf{2.0454} & \textbf{2.0431} &  &\textbf{1.8441} & \textbf{1.9360}\\
\bottomrule
\end{tabular}}
\end{minipage}
\end{center}
\end{table}

\begin{table}[h!]
\begin{center}
\caption{Percentage of simulations in which different candidate copula models are selected with the omnibus procedure.}
\label{tab:OMNISIM}
\begin{tabular}{@{}l l l l c c c c  @{}}
\toprule
\multirow{2}{*}{Scenario}  && \multirow{2}{*}{True copula}  &&  \multicolumn{4}{c}{Candidate model}\\ \cline{5-8}
 && &  &Clayton & Frank & Gumbel & Joe \\
\midrule
\multirow{2}{*}{No censoring}&& Frank &  & 0.876 &  0.322 & 0.853  & 0.949 \\
&&Joe &  & 0.897 & 0.680 & 0.845 & 0.578  \\
\midrule
\multirow{2}{*}{Single-censoring} && Frank && 0.880 & 0.274  & 0.884  & 0.962 \\
&&Joe && 0.892 & 0.844 &  0.868  & 0.396  \\
 \midrule
 \multirow{4}{*}{Double-censoring}  && Clayton && 0.276 & 0.983  & 0.929 & 0.822 \\
 && Frank && 0.865 & 0.392  & 0.947  & 0.796 \\
 && Gumbel && 0.906 & 0.992  & 0.370  & 0.732  \\
&&Joe && 0.936 &  0.942 & 0.962  & 0.160  \\
\bottomrule
\end{tabular}
\end{center}
\end{table}

\subsection{$L^2$-norm}
\label{sec:L2norm_sim}
Next, we perform a second study in which we simulate bivariate samples of $n=500$ observations from various copulas with Kendall's tau equal to $0.2$, $0.4$, and $0.6$. We use the same exponential distributions of parameters equal to one for the marginal distributions of $(T_1, T_2)$. We again consider three censoring schemes: one with no censoring, one where we allow only one variable to be censored, and one where both variables are subject to censoring. For the double-censoring scheme, we also analyze the independence case. 

When working with complete data, we use \cite{genest1993}'s estimator $\hat{K}_n(\nu)$ to compute $D(\hat{\alpha})$:
\begin{align*}
    \hat{K}_n(\nu) = \frac{1}{n}\#\{i \vert \nu_i \leq \nu \},
\end{align*}
with
\begin{align*}
    \nu_i = \frac{1}{n-1}\# \{(t_{1,(j)}, t_{2,(j)})  \vert  t_{1,(j)}<t_{1,(i)}, t_{2,(j)}<t_{2,(i)}\}.
\end{align*}

For the single-censoring scenario, we use a \cite{wang2000}'s estimator $\hat{K}_n(\nu)$ similar to Section~\ref{sec:estimator}, but with a different estimator for the joint distribution $\hat{F}(\vect{y})$. Namely, we compute the estimator proposed by \cite{akritas1994} for single-censoring scenarios. Assuming that $T_1$ is the variable that can be censored by $X_1$, we have:
\begin{align*}
    \hat{F}(y_1, t_2) &= \int_0^{t_2} \hat{F}_{Y_1 \vert T_2}(y_1 \vert z)d\hat{F}_{T_2}(z) \\
    &= \frac{1}{n} \sum_{k=1}^n \mathds{1}[0 \leq t_{2,k} \leq t_2]\hat{F}_{Y_1 \vert T_2}(y_1 \vert t_{2,k}),
\end{align*}
where $\hat{F}_{Y_1 \vert T_2}(y_1 \vert z)$ is \cite{beran1981}'s estimator from Equation~\eqref{eq:beran}.

For the two censoring scenarios, we simulate the censoring variables using exponential distributions with parameters such that at least $20\%$ of observations have one (or two) censored components. For each sample, we then perform the bootstrap procedure described in Section~\ref{sec:L2norm} to obtain estimates of the $L^2$-norm distance between $\hat{K}_n(\nu)$ estimated non-parametrically for the data and the corresponding $K_{\hat{\alpha}}(\nu)$ for different candidate models.

Table~\ref{tab:L2normSIM} presents the results of $1000$ simulations for each sample. Each value in this table corresponds to the pseudo $p$-value from Section~\ref{sec:L2norm} and should be understood as the percentage of bootstrap simulations in which the candidate model was not selected as the best model for the simulated sample. That is the percentage of simulations for which the $L^2$-norm distance between $\hat{K}_n(\nu)$ and $K_{\hat{\alpha}}(\nu)$ for the candidate model was not the smallest one compared to the other models under consideration. We use samples from both Frank and Joe copulas in the no-censoring and single-censoring schemes. At the same time, in the double-censoring scenario, we analyze samples from the Clayton, Frank, Gumbel, and Joe copulas. 

We observe that in all three scenarios for each copula, the pseudo $p$-value is the smallest for the candidate model corresponding to the true copula. For example, in the double-censoring scenario, when we simulate the data from a Frank copula with Kendall's tau equal to $0.4$, we observe that the Frank copula is rejected as an appropriate candidate model in just $1.5\%$ of the bootstrapped simulations performed. The Gumbel model is rejected in $98.5\%$ of simulations, and the Clayton and Joe copulas are rejected in all simulations. In addition, we observe that as the value of Kendall's tau increases, the percentage of rejection for the correct model decreases. For the Frank sample in the double-censoring scenario, $8.7\%$ of the bootstrap simulations reject the Frank model when $\tau = 0.2$. This percentage decreases to $1.5\%$ when $\tau = 0.4$ and to $0.9\%$ when $\tau = 0.6$. We observe similar results and trends in all scenarios for each copula considered. This showcases the strength of our test. 

\begin{figure}[h!]
\centering
\includegraphics[width=.6\textwidth]{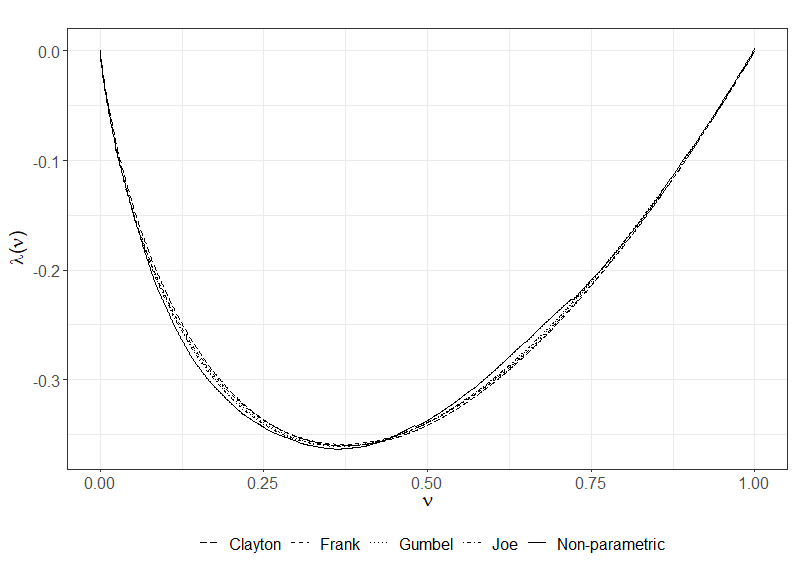}
\caption{Graphical comparison of the estimated $\hat{\lambda}(\nu)$ functions for the independent samples. }
\label{fig:IND_plot}
\end{figure}

In addition, for the doubly-censored scenario, we simulate samples from each of the four copulas with Kendall's tau equal to $0$. This is equivalent to simulating four bivariate samples from the independence copula. We observe that the results of the bootstrap simulations are much more random. For the Joe sample, for example, $45\%$ of simulations reject the Clayton model, all reject the Frank model, $80\%$ reject the Gumbel model, and $75\%$ reject the Joe model. This is to be expected since, when $\tau = 0$, all four copulas converge towards the same independence copula. The estimates that we thereby get for $K_{\hat{\alpha}}(\nu)$ and $\lambda_{\hat{\alpha}}(\nu)$ all converge towards the same functions. The choice of the smallest distance between $\hat{K}_n(\nu)$ and $K_{\hat{\alpha}}(\nu)$ for the different candidate models thus becomes much more random. This is well illustrated in Figure~\ref{fig:IND_plot} that shows the plot of $\hat{\lambda}(\nu)$ and those of $\lambda_{\hat{\alpha}}(\nu)$ for the four candidate models considered—the curves for the Clayton, Frank, Gumbel, and Joe copula overlap. 

\begin{table}[h!]
\begin{center}
\caption{Pseudo $p$-values for different candidate copula models under different censoring scenarios.}
\label{tab:L2normSIM}
\begin{tabular}{@{}l l l c c c c c  @{}}
\toprule
\multirow{2}{*}{Scenario}  && \multirow{2}{*}{True copula} & \multirow{2}{*}{$\tau$} &  \multicolumn{4}{c}{Candidate model}\\ \cline{5-8}
 && &  &  Clayton & Frank & Gumbel & Joe \\
\midrule
\multirow{6}{*}{No censoring} && \multirow{3}{*}{Frank} & 0.2 & 1.000 & 0.091 & 0.909 & 1.000 \\
&& & 0.4 & 1.000 &  0.067 & 0.948  & 0.985 \\
&& & 0.6 & 1.000 & 0.001 & 0.999  & 1.000 \\ \cline{3-8}
&&\multirow{3}{*}{Joe} & 0.2 & 1.000 & 0.999 & 0.924  &  0.077  \\
&& & 0.4 & 1.000 & 1.000 & 0.998 & 0.002  \\
&& & 0.6&  1.000& 1.000 & 0.999 & 0.001  \\
\midrule
\multirow{6}{*}{single-censoring} && \multirow{3}{*}{Frank} & 0.2 & 1.000 & 0.078 & 0.923 & 0.999  \\
&& & 0.4 & 1.000 & 0.054  & 0.946  & 1.000 \\
&& & 0.6 & 1.000 & 0.012 &  0.999 & 0.989 \\ \cline{3-8}
&&\multirow{3}{*}{Joe} & 0.2 & 1.000 & 1.000 & 0.928  &  0.072  \\
&& & 0.4 & 1.000 & 1.000 &  0.987  & 0.013  \\
 &&& 0.6 & 1.000 & 1.000 & 0.999   & 0.001 \\
 \midrule
 \multirow{15}{*}{double-censoring}  && \multirow{4}{*}{Clayton} & 0.0 & 0.470 & 0.890 & 0.790 & 0.850  \\
  &&& 0.2 & 0.170 & 0.831 & 0.999 & 1.000 \\
 &&& 0.4 & 0.079 & 0.921  & 1.000 & 1.000 \\
 &&& 0.6 & 0.000 & 1.000 & 1.000 & 1.000 \\ \cline{3-8}
 && \multirow{4}{*}{Frank} & 0.0 & 0.230 & 1.000 & 0.770 & 1.000  \\
  &&& 0.2 & 0.986 & 0.087 & 0.928 & 0.999  \\
 &&& 0.4 & 1.000 & 0.015  & 0.985  & 1.000 \\
 &&& 0.6 & 1.000 & 0.009 & 0.991 & 1.000 \\ \cline{3-8}
 && \multirow{4}{*}{Gumbel} & 0.0 & 0.620 & 1.000  & 1.000  & 0.038   \\
  &&& 0.2 & 1.000 &  0.970 &  0.168 &  0.862 \\
 &&& 0.4 & 1.000 &  0.975 &  0.135 &  0.890 \\
 &&& 0.6 & 1.000 & 0.986 & 0.057 & 0.957 \\ \cline{3-8}
&&\multirow{4}{*}{Joe} & 0.0 & 0.450 & 1.000 & 0.800  &  0.750 \\
 &&& 0.2 & 1.000 & 1.000 & 0.843  &  0.157  \\
 &&& 0.4 & 1.000 & 1.000 &  0.929  & 0.071  \\
 &&& 0.6 & 1.000 & 1.000 &  0.949  & 0.051 \\
\bottomrule
\end{tabular}
\end{center}
\end{table}

\subsection{Goodness-of-fit test}
\label{sec:wang_sim}
Using the same censoring scenarios, marginal distributions, and copulas as in Section~\ref{sec:L2norm_sim} with Kendall's tau equal to $0.2$, $0.4$, and $0.6$, we now apply \cite{wang2010}'s goodness-of-fit test described in Section~\ref{sec:wang}. We present the results of 1000 simulations for simulated samples of size $n=200$ in Table~\ref{tab:tau}. For the single-censoring and double-censoring scenarios, we use $M=5$ imputed datasets. 

\begin{figure}[h!]
\centering
\includegraphics[width=.6\textwidth]{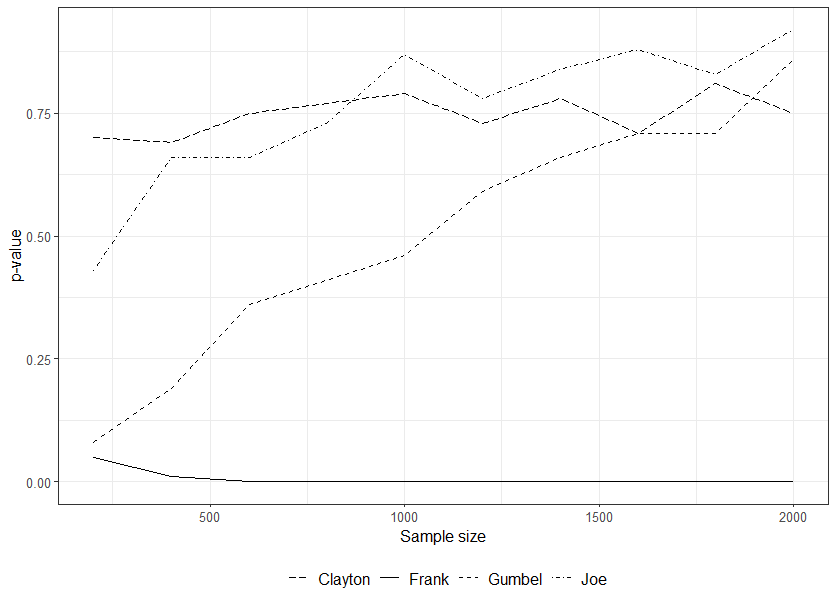}
\caption{Evolution of the $p$-values of competing models based on $1000$ simulations for varying sample sizes simulated from the Frank copula with Kendall's tau equal to $0.4$ in a no censoring scenario.}
\label{fig:sample_size}
\end{figure}

We observe similar results to those obtained by \cite{wang2010} and to those from Section \ref{sec:L2norm_sim}. When dependence is weak, the test does not always accurately predict the best model for the data. In particular, we see that in the single-censoring scenario when the sample comes from the Frank copula with $\tau = 0.2$, even though the lowest percentage of rejection of the null hypothesis is correctly observed when the candidate model is the Frank copula, this percentage is also relatively low for the three alternative models. As already observed in \cite{wang2010}, the power of the test, however, increases when we increase the sample size. To illustrate this, we present in Figure~\ref{fig:sample_size} what happens to the $p$-values in the no-censoring scenario when the true model is the Frank copula with $\tau = 0.4$ when we progressively increase the sample size from $n=200$ to $n=2000$. We work with increasing steps of $200$ observations; in each case, we perform 1000 simulations. The strength of the test increases quite steeply with the sample size. For the Frank candidate model, the resulting $p$-value starts at a low level of approximately $5\%$, as was reported in Table~\ref{tab:tau}, and quickly tends to $0$ as we increase the sample size. The Gumbel copula displays a low $p$-value of less than $10\%$ for the smallest sample size. However, it greatly increases such that for a sample of $2000$ observations, it is correctly rejected by over $85\%$ of simulations. The Joe and Clayton candidate models display similar increasing trends, even though they started with a higher $p$-value for small sample sizes than the Gumbel copula.

Similarly to the results of the $L^2$-norm in Table~\ref{tab:L2normSIM}, we also observe in Table~\ref{tab:tau} that the strength of the test increases with the level of dependence. In all scenarios and for all copula models, the percentage of rejection of the null hypothesis increases with $\tau$ when the null hypothesis is incorrectly specified and decreases with $\tau$ when the true and null copula coincide. 

\begin{table}[h!]
\begin{center}
\caption{Percentage of rejection of the null hypothesis for different copulas $n=200$}
\label{tab:tau}
\begin{tabular}{@{}l l l c c c c c @{}}
\toprule
\multirow{2}{*}{Scenario}&&\multirow{2}{*}{True copula} & \multirow{2}{*}{$\tau$} &  \multicolumn{4}{c}{Copula under $H_0$}\\ \cline{5-8}
&& &  &  Clayton & Frank & Gumbel & Joe \\
\midrule
\multirow{6}{*}{No censoring} &&\multirow{3}{*}{Frank} & 0.2 & 0.351 & 0.082 &  0.030 & 0.074 \\
&& & 0.4 &  0.728  & 0.058 &  0.086& 0.679\\
&& & 0.6 & 0.859  & 0.022 & 0.196 & 0.956\\ \cline{3-8}
&&\multirow{3}{*}{Joe} & 0.2 & 0.814 & 0.206 & 0.248 & 0.022\\
&& & 0.4 &  0.922 & 0.776 & 0.724 & 0.018 \\
&& & 0.6 & 0.944  & 0.970 & 0.942 & 0.003 \\
\midrule
\multirow{6}{*}{single-censoring} &&\multirow{3}{*}{Frank} & 0.2 & 0.142 & 0.032 & 0.229  & 0.142\\
&& & 0.4 &  0.712  & 0.084 & 0.800 & 1.000\\
&& & 0.6 &  0.760 & 0.000& 0.968 & 0.990\\ \cline{3-8}
&&\multirow{3}{*}{Joe} & 0.2 & 0.720 & 0.030 & 0.052 & 0.016\\
&& & 0.4 & 0.970  & 0.232 & 0.081 & 0.007 \\
&& & 0.6 & 0.980  & 0.400 & 0.131 & 0.004 \\
\midrule
\multirow{12}{*}{double-censoring} &&\multirow{3}{*}{Clayton} & 0.2 & 0.097 & 0.031 & 0.512  & 0.239\\
&& & 0.4 &  0.021  & 0.203 & 0.835 & 0.945 \\
&& & 0.6 & 0.016  & 0.645 & 0.968 & 0.958\\ \cline{3-8}
&&\multirow{3}{*}{Frank} & 0.2 & 0.391 & 0.022 & 0.057  & 0.029 \\
&& & 0.4 & 0.810   & 0.012 & 0.850 & 0.999\\
&& & 0.6 & 0.944  & 0.001 & 0.975 & 1.000\\ \cline{3-8}
&&\multirow{3}{*}{Gumbel} & 0.2 &  0.456 & 0.082 & 0.010 & 0.012\\
&& & 0.4 &  0.848 & 0.172 & 0.030 & 0.051 \\
&& & 0.6 & 0.962  & 0.458 & 0.025 & 0.775\\ \cline{3-8}
&&\multirow{3}{*}{Joe} & 0.2 & 0.990 & 0.468 & 0.535 &0.081\\
&& & 0.4 &  1.000 & 0.530 & 0.727 & 0.002 \\
&& & 0.6 & 1.000  & 0.880 & 0.843 & 0.000 \\
\bottomrule
\end{tabular}
\end{center}
\end{table}

\subsection{Impact of the limit}
In Section~\ref{sec:notation}, we introduced a limit $\omega_i$ for $i=1,2$ acting as an additional censoring layer. Depending on the context, these limits can be imposed by practitioners, authorities, etc. They can be the maximum time before treatment is considered ineffective, even if the patient is still alive, or the maximum payment delay authorized for a claim for example.

In this section, we conduct a simulation study to assess the impact of $\omega$ on the strength and form of dependence between two censored variables. 

Consider a bivariate vector of time-to-events $\vect{T} = (T_1, T_2)$ with log-normal distributions such that $\mu_{T_1} = 8$, $\mu_{T_2} = 7$, $\sigma_{T_1} = 1$ and $\sigma_{T_2} = 3$. We set the correlation $\rho$ between $T_1$ and $T_2$ at $0.35$. Let $\vect{X}=(X_1, X_2)$ be the censoring vector with $X_1$ independent from $X_2$. Both $X_1$ and $X_2$ are log-normally distributed, with parameters $\mu_{X_1}$ and $\mu_{X_2}$ chosen to allow for different levels of censorship. We also set $\sigma_{X_1} = 1$ and $\sigma_{X_2}=1$. We perform three simulation studies: one with low levels of censorship for both variables (approx. $2\%$ and $1\%$ for, respectively, $T_1$ and $T_2$), one with medium levels of censoring (approx. $25\%$ and $40\%$) and one with higher levels of censorship (approx. $65\%$ and $75\%$ for, respectively, $T_1$ and $T_2$). Finally, consider the limits $\omega_1$ and $\omega_2$ such that we observe the bivariate vector $\vect{Y}=(Y_1, Y_2) = \big(\min(T_1, X_1, \omega_1),\min(T_2,X_2,\omega_2)\big)$.

We use the strategy laid out in Section~\ref{sec:Model} to derive the joint distribution \eqref{eq:joint} from Beran's estimators. As in \cite{akritas2003}, we choose the weights $w(\vect{y}) = 0.5$ for the sake of simplicity although better results could be obtained by optimizing them. Using the estimated joint distribution, we then derive Kendall's tau and estimate $\hat{\lambda}_n(\nu)$ for a sample of $n=500$ observations. We use different quantiles from the distributions of $Y_1$ and $Y_2$ to set the values of the limits $\omega_1$ and $\omega_2$.

\begin{figure}[h!]
\centering
\includegraphics[width=.9\textwidth]{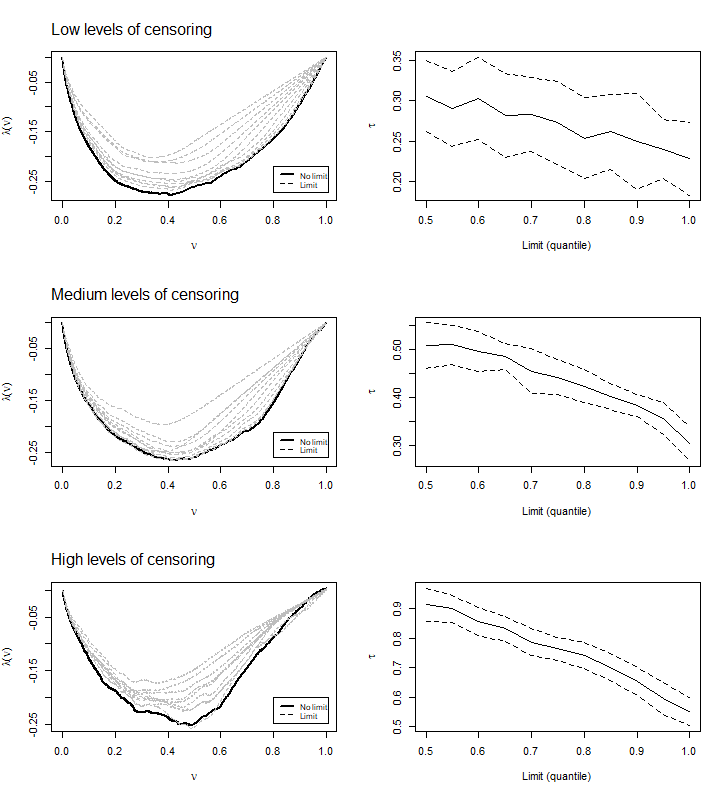}
\caption{Plots of $\lambda(\nu)$ and Kendall's tau for the joint distribution using different limits and different levels of censoring. Low, medium and high levels of censoring correspond to, approximately, $5\%$, $30\%$ and $75\%$.}
\label{fig:Fig5}
\end{figure}

Figure~\ref{fig:Fig5} displays the results of decreasing the limit as well as the level of censoring on the form of dependence via the function $\hat{\lambda}_n(\nu)$ and the strength of dependence via Kendall's tau. We observe that neither the limit nor the level of censoring seem to have an impact on the form of dependence: in all cases, the shape of the generator function remains similar. As we increase the level of censoring and the limit, we notice that the shape of dependence becomes even more pronounced. This is in line with the results from Sections \ref{sec:L2norm_sim} and \ref{sec:wang_sim} where we noted that, even though both tests were able to correctly choose the most appropriate copula for a given sample and low dependence levels, their strength increased with Kendall's tau. 

\section{Application to automobile insurance claims}
\label{sec:application}
In this section, we apply our proposed non-parametric approach to a recent Canadian automobile insurance dataset that includes over $600,000$ claims that occurred between January 2015 and June 2021. Each policy in force in this dataset provides four different coverages to its holder, and each claim can fall under one or more of these coverages. 

We focus on the Accident Benefits and Bodily Injury coverages, which are the most important cost-wise in the portfolio. Our goal is to model the dependence between them using their activation delays. These are defined as the time elapsed between the reporting date of the claim and the date at which the insurer first triggers the coverage in the claims management system. This is the date the insurer records that the claim falls under that specific coverage. Focusing on the bivariate case, i.e., a policy providing two coverages, we denote the vector of activation delays by $\vect{T}=(T_1, T_2)$. If the claim settles before one of the coverages is triggered, then the delay for this coverage is censored. The censoring variable is the claim settlement delay we denote by $\vect{X}=(X_1, X_2)$. Note that we have $X_1 = X_2$ in this specific application since we assume that a claim only settles once. In addition, policyholders may be unable to receive compensation for a specific coverage if a certain amount of time has passed since reporting. This limit on the delays can depend on the nature of the claims, local or governmental regulations, or even on company-level rules of the insurer. Both internal and external agents can set it. This is our limit $\omega_i$, for $i=1,2$ introduced in Section~\ref{sec:notation}. Suppose a coverage has not been triggered yet when this limit passes. In that case, its activation delay becomes censored, even if the claim remains open.

\begin{figure}[h!]
\centering
\includegraphics[width=1\textwidth]{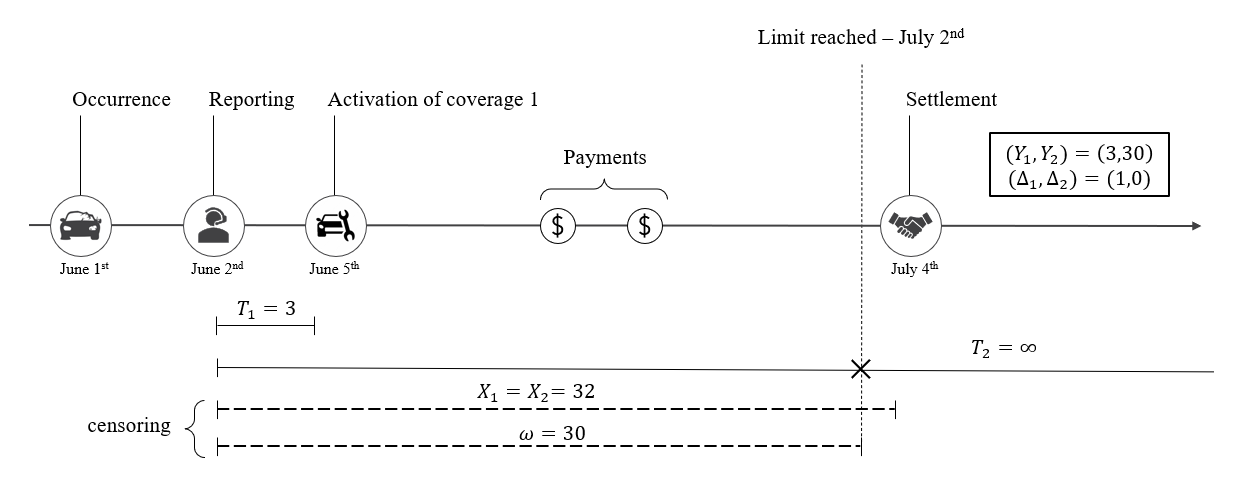}
\caption{Typical claim development and illustration of the notation used.}
\label{fig:schema}
\end{figure}

Figure~\ref{fig:schema} illustrates the typical development of a claim and the notation introduced above. In this example, a claim occurred on June $1^{\text{st}}$ and was reported the following day. On June $5^{\text{th}}$, the insurer records that the claim has impacted coverage 1. Recording the delays in days, the activation delay for this coverage is then $T_1 = 3$. Assuming a limit of $\omega_1 = \omega_2 = 30$ days on the activation delays, coverage 2 becomes censored on July $2^{\text{nd}}$, even if the claim is still open. It settles two days later, leading to $X_1 = X_2 = 32$. The observed delays are then given by $(Y_1, Y_2) = \big(\min(T_1, X_1, \omega_1), \min(T_2, X_2, \omega_2)\big) = \big(\min(3,32,30), \min(\infty, 32, 30)\big)=(3,30)$.

Considering the observed activation delays shown (in periods of six months) in Table~\ref{tab:actidelays} for the Accident Benefits and Bodily Injury coverages in the portfolio at hand, we set the limit at $\omega_1 = \omega_2 =730$ days, equivalent to two calendar years. Almost $94\%$ of all Accident Benefits claims trigger that coverage within six months of reporting, and most of the remaining claims will be labelled as Accident Benefits no later than a year after the reporting date. The activation delays are longer for Bodily Injury claims, with $1.69\%$ of them taking two years or more after reporting to be labelled as such.

\begin{table}[h]
\begin{center}
\caption{Percentage of claims with different activation delays, shown in periods of six months.}
\label{tab:actidelays}
\begin{tabular}{@{}l r r r r r@{}}
\toprule
& \multicolumn{5}{c}{Activation delays} \\
\midrule
  Coverage & \multicolumn{1}{c}{No delay} & \multicolumn{1}{c}{1 period} & \multicolumn{1}{c}{2 periods } & \multicolumn{1}{c}{3 periods } & \multicolumn{1}{c}{$\geq 4$ periods } \\ 
 \midrule
 Accident Benefits  & 93.84 & 5.73 & 0.29 & 0.08 & 0.06\\
 Bodily Injury  & 85.86 & 9.86 & 1.46 & 1.13 & 1.69\\
 \bottomrule
\end{tabular}
\end{center}
\end{table}

\begin{figure}[h!]
\centering
\includegraphics[width=1\textwidth]{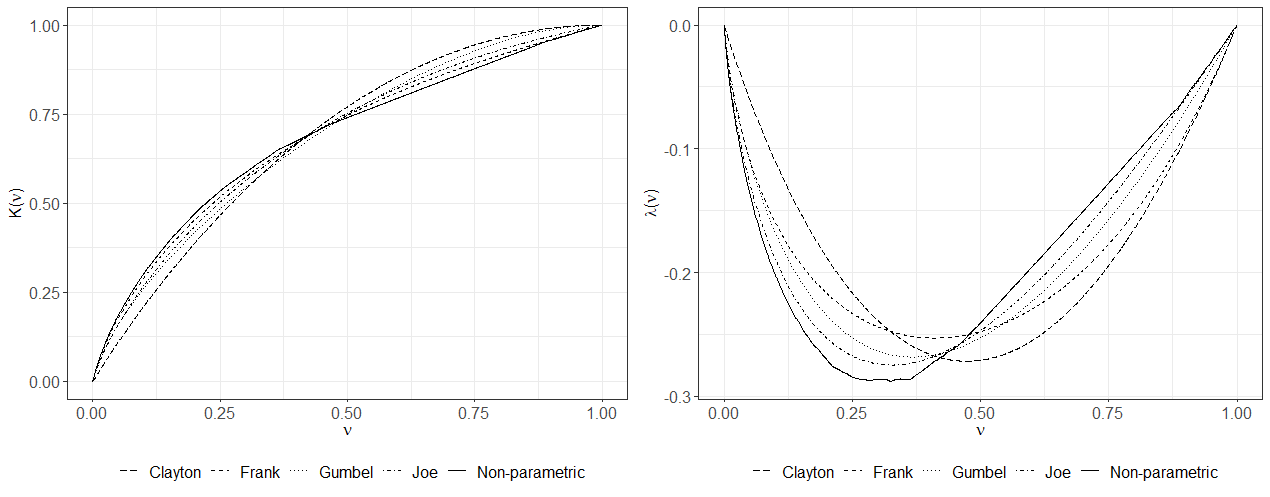}
\caption{$K(\nu)$ and $\lambda(\nu)$ for the different copulas and for the Canadian automobile insurance dataset with a limit at 730 days.}
\label{fig:Fig1}
\end{figure}

We first apply the approach described in Section~\ref{sec:Model} to graphically find the copula best fitted to the dataset at hand. We present the plots of the non-parametric estimators $\hat{K}_n(\nu)$ and $\hat{\lambda}_n(\nu)$ in Figure~\ref{fig:Fig1}, along with the corresponding fitted curves for four candidate Archimedean models. In both plots, the non-parametric estimator is closest to the curves of the Joe copula. With an estimated $\hat{\tau} = 0.2705$, the Joe copula with $\hat{\alpha} = 1.6652$ best fits our data. 

The omnibus procedure first confirms this result. We estimate the dependence parameters for the competing models using the likelihood in Equation~\eqref{eq:lik}. The results shown in the third and fourth columns of Table~\ref{tab:validation} indicate that the Joe copula is most appropriate since it presents the smallest difference between $\hat{\alpha}$ and $\hat{\alpha}^*$. Similarly, the results of 1000 bootstrapped simulations for the $L^2$-norm validation approach shown in the fifth column of Table~\ref{tab:validation} show that $D(\hat{\alpha})$ is the smallest for the Joe copula. We finally apply \cite{wang2010}'s goodness-of-fit test and find again in the last column of Table~\ref{tab:validation} that the Joe model presents the smallest percentage of rejection when we perform 1000 simulations for each competing model.

\begin{table}[h]
\begin{center}
\caption{Results validation, based on 1000 bootstrapped simulations for the $L^2$-norm and 1000 simulations for \cite{wang2010}'s test.}
\label{tab:validation}
\begin{tabular}{@{}l c c c c c @{}}
\toprule
\multirow{2}{*}{Copula} & \multirow{2}{*}{$\hat{\tau}$} &  \multicolumn{2}{c}{Omnibus procedure} & $L^2$-norm & \cite{wang2010}\\
 &  &  $\hat{\alpha}^*$ & $\hat{\alpha}$ & $D(\hat{\alpha})$ & $\%$ rejection $H_0$\\
\midrule
Clayton & \multirow{4}{*}{0.2705} & 0.2432 & 0.7417 & 0.00399& 0.9994\\
Frank &  & 1.1166 & 2.5612 & 0.00180& 0.9265\\
Gumbel &  & 1.0554 & 1.3708& 0.00084& 0.8742\\
Joe &  & 1.3821 & 1.6652& 0.00033& 0.0476\\
\bottomrule
\end{tabular}
\end{center}
\end{table} 

\begin{figure}[h!]
\centering
\includegraphics[width=.9\textwidth]{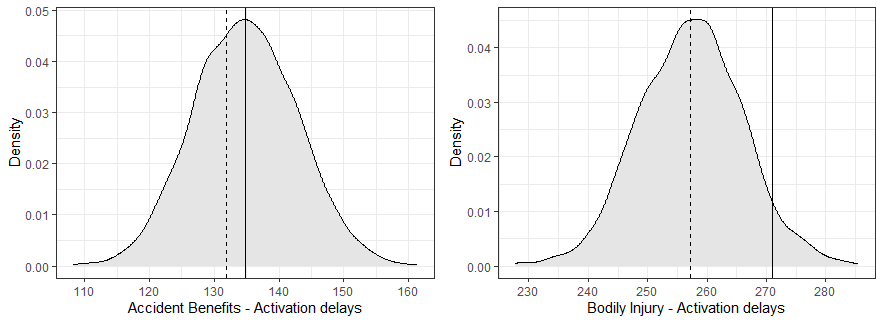}
\caption{Densities of the simulated activation delays for the Accident Benefits and Bodily Injury coverages, taking the $1^{\text{st}}$ of January 2019 as valuation date.}
\label{fig:Fig_delays}
\end{figure}

Taking the resulting Joe copula as best fitted to our data, we now show the results of some simulations for the activation delays, taking the $1^\text{st}$ of January 2019 as the valuation date. Figure~\ref{fig:Fig_delays} shows the simulated densities for the delays of both coverages. The dotted lines in both plots represent the average of the simulations, while the continuous line shows the true mean delays. We observe that the simulations provide a good match for the observed data. The simulated mean activation delay for the Accident Benefits coverage is 134.79 days compared to an observed average delay of 131.52. The average simulated delay for the Bodily Injury claims is 257.26 days, and the observed average is slightly higher at 271.08 days. This larger difference in the predictions for the Bodily Injury claims stems from the greater variability typically observed for this coverage. 

When a claim occurs, we can thus predict the activation delays for both coverages with a rather good level of accuracy. This can then be incorporated into a larger claims reserving model and enhance the predictions for the total portfolio reserve. 

\section{Conclusion}
\label{sec:conclusion}
In this paper, we present a non-parametric estimator for the generator function of the Archimedean family of copulas and extend it to flexible censoring scenarios. Following the initial idea of \cite{genest1993}, we propose a simple graphical selection procedure that works well under various levels of censoring. We then present three approaches to validate the results, all applicable to flexible censoring schemes for the Archimedean family. 

We illustrate both the non-parametric approach and validation methods using various simulation studies. We discuss the impact of a limit we introduce as a particular type of censoring and show that it does not affect the model selection. We then apply this methodology to a real automobile insurance dataset using activation delays for two insurance coverages provided within each policy in force. We show good results when using real, large datasets and can easily retrieve the most appropriate copula model. We further show that the copula model resulting from the non-parametric approach provides accurate estimates of the activation delays for both coverages. 

The approach presented in this paper could be used as part, for example, of a claims reserving model such as that proposed in \cite{michaelides2023} or \cite{cote2022} where the estimation of the activation delays can replace, respectively, the activation patterns and the multinomial estimation of the claim type. A similar methodology could also be used in different models such as that proposed in \cite{antonio2014} where, for example, the next event's exact time for reported but not settled claims could be extended to a multivariate framework by considering simultaneously, as in Section \ref{sec:application}, different business lines or insurance coverages. Alternative dependence models could also be considered, particularly other families of copulas, as in \cite{lopez2015}.







\newpage
\bibliographystyle{unsrt}  
\bibliography{main}  

\newpage
\appendix

\section{Distributions of $U$ and $V$ proposed in \cite{wang2010} for diverse censoring patterns} 
\label{app:1}
The following results are based on Corollary 1., Corollary 2. and Theorem 4. from \cite{wang2010}.

Let $(T_1,T_2)$ be a bivariate vector submitted to censoring by the bivariate continuous vector $(X_1,X_2)$, as described in Section~\ref{sec:notation}. Let $S_1(t_1)$ and $S_2(t_2)$ be the marginal survival functions of, respectively, $T_1$ and $T_2$, whose dependence function can be modelled by an absolutely continuous Archimedean copula such that
\begin{align*}
    C(S_1(t_1),S_2(t_2)) = S(t_1,t_2) = \phi^{-1}\Big\{\phi(S_1(t_1)) + \phi(S_2(t_2))\Big\}.
\end{align*}
Then, we have that:

\begin{enumerate}
    \item if both observations are censored, the distribution function of $(V \vert T_1>x_1, T_2> x_2)$ is
    \begin{align*}
        F_1(v,x_1,x_2) = \frac{1}{S(x_1,x_2)}\Bigg[ v - \frac{\phi(v) - \phi\{S(x_1,x_2)\}}{\phi'(v)} \Bigg], \hspace{0.5cm} 0\leq v \leq S(x_1,x_2)
    \end{align*}
    and the distribution function of $(U \vert T_1 > x_1, T_2 > x_2)$ is uniformly distributed on the interval 
    \begin{align*}
        \Bigg[ \frac{\phi\{S_1(x_1)\}}{\phi(v)}, 1 - \frac{\phi\{S_2(x_2)\}}{\phi(v)} \Bigg];
    \end{align*}
    \item if only the second observation is censored, the distribution function of $(V \vert T_1 = t_1, T_2 > x_2)$ is
    \begin{align*}
        F_2(v,t_1,c_2) = \frac{p'(\phi(v))}{p'(\phi(S(t_1,x_2)))}, \hspace{0.5cm} 0\leq v \leq S(t_1,x_2)
    \end{align*}
    with $p(.) = \phi^{-1}(.)$ and the distribution function of $(U \vert T_1 = t_1, T_2 > x_2)$ is
    \begin{align*}
        G_2(y,t_1,x_2) = \frac{p'\{ \phi(S_1(t_1))/u \}}{p'\{ \phi(S(t_1,x_2))\}}, \hspace{0.5cm} 0 \leq u \leq \frac{\phi\{ S_1(t_1) \}}{\phi\{ S(t_1,x_2) \}};
    \end{align*}
    \item if only the first observation if censored, the distribution function of $(V \vert T_1 > x_1, T_2 = x_2)$ is
    \begin{align*}
        F_3(v,x_1,t_2) = \frac{p'(\phi(v))}{p'(\phi(S(x_1,t_2)))}, \hspace{0.5cm} 0\leq v \leq S(x_1,t_2)
    \end{align*}
    and the distribution function of $(U \vert T_1>x_1, T_2 = t_2)$ is
    \begin{align*}
        G_3(y,x_1,t_2) = \frac{p'\{ \phi(S_2(t_2))/(1-u) \}}{p'\{ \phi(S(x_1,t_2))\}}, \hspace{0.5cm} \frac{\phi\{ S_1(x_1) \}}{\phi\{ S(x_1,t_2) \}} \leq u \leq 1;
    \end{align*}
    \item if both observations are uncensored, the distribution function of $(V  \vert  T_1 = t_1, T_2 = t_2)$ is the same as in Proposition 1.1. from \cite{genest1993} and the conditional distribution of $(U \vert V=v)$ is just a uniform distribution on $[0,1]$ independent of $V$.
\end{enumerate}



\end{document}